\documentstyle{res}
\input epsf

\begin{document}
\def\lya{{\rm Ly}$\alpha$}
\def\ha {{\rm H}$\alpha$}
\def\han{\ha}
\def\hb {{\rm H}$\beta$}
\def\arcsecs{\ifmmode {''}\else $''$\fi}
\def\eg{{\it e.g.}}
\def\et{{\it et al.}\ }
\def\cf{{\it c.f.}}
\def\ie{{\it i.e.}}
\def\solar{\ifmmode_{\mathord\odot}\else$_{\mathord\odot}$\fi}
\def\Msun{\M$_\odot$}
\def\mic{~$\mu$m}
\def\kp{{\rm K}$^{\prime}$}

\def\smalltext#1{
\noindent{\small #1}
\baselineskip=14pt 
}
\def\spose#1{\hbox to 0pt{#1\hss}}
\def\simlt{\mathrel{\spose{\lower 3pt\hbox{$\mathchar"218$}}
     \raise 2.0pt\hbox{$\mathchar"13C$}}}
\def\simgt{\mathrel{\spose{\lower 3pt\hbox{$\mathchar"218$}}
     \raise 2.0pt\hbox{$\mathchar"13E$}}}
\def\simpropto{\mathrel{\spose{\lower 3pt\hbox{$\mathchar"218$}}
     \raise 2.0pt\hbox{$\propto$}}}


\def\thesection {\arabic{section}}
\def\thefigure{\arabic{figure}}

     
\def\thebibliography#1{\section*{References}\baselineskip 13pt \parskip 0pt
\list
 {\arabic{enumi}.}{\settowidth\labelwidth{[#1]}\leftmargin\labelwidth
 \advance\leftmargin\labelsep
 \usecounter{enumi}}
 \def\newblock{\hskip .11em plus .33em minus -.07em}
 \sloppy
 \sfcode`\.=1000\relax 
}
\let\endthebibliography=\endlist

\title{
Our Second Look at the Immature Universe:  \\
The Infrared View}

\author{Matthew Malkan \\
{\it Division of Astronomy, University of California, Los Angeles}}

\maketitle

\section*{Abstract}
\begin{quotation}
\footnotesize\noindent
The long-awaited promise of studying high-redshift galaxies at long
wavelengths has been partially eclipsed by progress at optical wavelengths,
mostly because of the number of available pixels. 
It is nonetheless essential to study optically selected high-redshift
galaxies at wavelengths longward of 1\mic, for several reasons.
One of the indications from such studies to date is that interstellar
dust (as an absorber of UV photons and re-emitter of IR photons) plays
an energetically significant role at high redshifts, just as it does
in present-day star-forming galaxies. 
This also provides a strong motivation for new searches which detect
high-redshift galaxies based on IR observations.  
Some of these are already succeeding in clarifying our view of galaxy
evolution in the Immature Universe.
\end{quotation}

\section{Introduction}

Optical observations jumped out to a rapid lead in studies of high-redshift (z$\ge$1)
galaxies, because of the sensitivity of large-format detectors, and the
accessibility of very strong spectral features in the far-UV rest frame
at z$\ge$ 2.5 (\lya, the trinity of deep interstellar absorptions due to 
Si~II, O~I, and Si~II,
at 1260, 1303, and 1334\AA, respectively,
and of course the dramatic Lyman continuum ``break"
which is robustly produced by both intrinsic and extrinsic absorption in
(presumably) all normal galaxies at z$\ge$2.5--3. (Steidel,\et\ 1996)
\footnote {As long as the intrinsic depth of the Lyman limit break is ``deep
enough" (e.g. two magnitudes), it can be used as an excellent indicator
of high-redshift galaxies in broad-band photometry which extends to short enough
wavelengths. However, the full depth of the Lyman break has still not been actually
measured in galaxies lacking active nuclei.  HUT provided a few strong upper
limits to the amount of Lyman continuum which manages to leak out of nearby
galaxies, and a deep FUV imaging program with the STIS MAMA detector in HST's
Cycle 9 should be
able to {\it detect} this emission in a sample of moderate-redshift ($1.1 \ge z \ge 1.4$)
starburst galaxies (GO program 8561).}
In surprisingly short order, this lead to an impressive list of results 
e.g. luminosity functions and integrated star formation rates, morphologies,
and even spatial clustering (Steidel, \et\ 1998).

It is certainly exciting to be able to see much data on an evolutionary 
phase in normal galaxies that was hardly observed at all less than ten years ago.
\footnote {The last time I attended a meeting at this lovely conference center,
in 1988, there were virtually {\it no} observational 
data on normal galaxies above redshifts of 1.}
However, as the initial euphoria of discovery has subsided, researchers have 
started to face the practical reality: all of this new information, as invaluable as
it is, rests fundamentally on the strong stellar continuum escaping from
these galaxies in the far-UV. It is well known that hardly any other
region of the electromagnetic spectrum is more sensitive to absorption by 
interstellar dust grains.
And this effect is especially strong when we are observing newly-formed stars,
since they are too short-lived to escape from  
the interstellar matter out of which they are born.
This leads immediately to the outstanding questions about these ``Lyman Break" galaxies: 
1) how can we estimate their properties, such as mass and metallicity,
2) which of their measured properties need to be corrected, 
3) by how much, 
and 4) what galaxies are missing altogether from these searches?
Thus the next task is to use observations at longer wavelengths to
obtain some answers.  This review is an attempt to see how long-wavelength
observations are starting to catch up with the optical ones, and fill in key gaps. 

\section{Long-Wavelength Observations of Galaxies Selected at Short Wavelengths}

The easiest way to build on optical studies of galaxies selected by their
rest-frame far-UV \footnote
{Most of these considerations also apply to the smaller number of high-redshift
galaxies which have been selected by their unusually strong \lya\ line emission.}
emission is to observe their rest-frame optical, infrared and radio emission.

The optical rest frame contains most of the strongest emission lines--our
most powerful diagnostics of the composition and kinematics of the ionized gas
associated with young stars.
An excellent example is near-IR spectroscopy, which was first accomplished with
UKIRT (Pettini \et\ 1998), 
and more recently with Keck (Teplitz \et\ 2000; Pettini \et\ 2000).
As an example of the new information available 
from strong, well-studied diagnostics in the
rest-frame optical is shown in Figure 1.
Such measurements of the [OIII] 5007 emission line 
in 5 LBGs indicate that the metallicities are mildly sub-solar 
(a half to a third--see Figure 2).
In addition to being useful for abundance estimates, the
rest-frame optical lines can measure galaxy dynamics
and masses.

\begin{figure}[t]
\begin{center}
\epsfxsize=\textwidth\epsffile{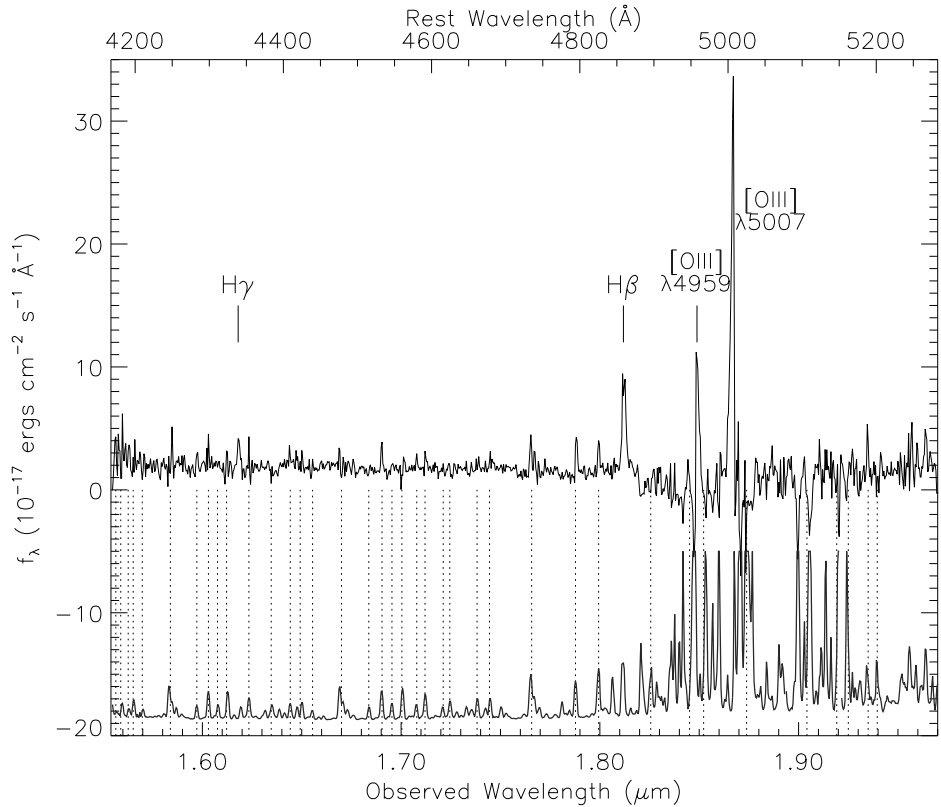}
\end{center}
\vspace{2.75truecm}
\caption{$\!\!$: NIRSPEC Spectrum of a z=2.7 Galaxy.}
\smalltext{H-band spectrum of MS1512-cB58.  
30 minutes of integration time on Keck.  The 1$\sigma$~errors
  are truncated and shifted downward on the plot.
  The ``break'' in the continuum at 1.82\mic~is a residual from the
  atmospheric absorption correction, not a drop in the galaxy's SED.}
\end{figure}

\begin{figure}[t]
\begin{center}
\epsfxsize=\textwidth\epsffile{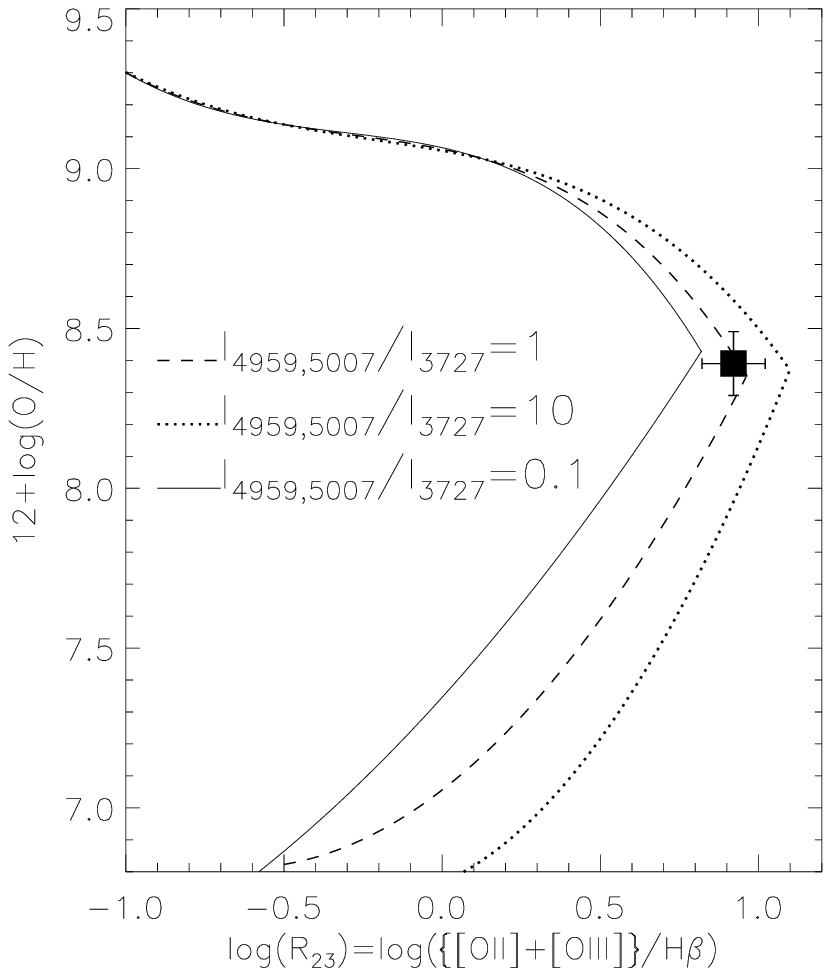}
\end{center}
\vspace{1.4truecm}
\caption{Emission line ratio $\log(R_{23}) = \log [I_{3727} + I_{5007}]/{\rm H}\beta$,
as a function of Oxygen abundance.}
\smalltext{The datapoint is for a z=3.4 galaxy
from Teplitz \et\ (2000). The curves show the range of extreme plausible
ratios for [OII]/[OIII].  
The metallicity is robustly constrained to be a half to a third solar.}
\end{figure}


Longer wavelength continuum observations (at 3--15\mic) 
constrain the old stellar population.  In the lensed z=2.7 
galaxy MS1512-cB58 (see its H-band spectrum in Figure 1), 
these ground-based and ISOCAM observations require that much
of the star formation occurred relatively shortly before we actually are observing
the galaxy (Bechtold \et\ 2000; Malkan \et\ 2000a).

Long-wavelength observations also provide 
estimates of the average extinction (through H recombination
line ratios and, less directly, the optical emission line-to-far-UV continuum
ratio)
which are in rough accord with more uncertain 
estimates from the far-UV continuum colors.
The tentative conclusion from these observations is that the far-UV emission from LBGs
needs very large corrections for dust absorption.
The {\it average} correction is ``only" a factor of 3--5,
while the total correction in integrated quantities such as global star formation
rates is higher, perhaps 7 times, because the more luminous galaxies appear to
be systematically redder and dustier (Pettini \et, 1998).  

The size of these corrections raises the concern that dustier galaxies are
missing altogether from the UV-selected samples. 
When infrared observers attempt to estimate global star formation rates
at redshifts of 1--2.5, these tend to be higher 
than those from the LBGs above z=2.5, even after large extinction corrections
(e.g. Rowan-Robinson \et\ 1997).
Of course it is possible that the integrated star formation in the Universe
really did peak at $z \sim 2$, but the lack of direct overlap between the
optical and infrared search methods has always worried me. 
The best way to settle this question will be deep observations at longer wavelengths.

\section{Selection and Discovery of High-Redshift Galaxies with Long-wavelength Observations}

We can do even more if we do not rely exclusively on optical observations to select
the galaxies.
The most conservative improvement to the optical observations is to include near-infrared
(e.g. J-, H- and possibly K-band) magnitudes in the photometric redshift methods, and this
is absolutely essential for $z \le 2.5$ where the Lyman
break feature is not accessible to ground-based observations, or even 
extremely long WFPC2 integrations in the 303W filter.
Examples of this approach are the Hubble Deep Fields (HDF-North,
Thompson \et\ 1999; and  
HDF-3516, Colbert, \et, 2000)

Nonetheless, these observations cannot answer the fourth and most haunting
of the questions posed above.
The second, more difficult approach needed to find out ``What is Missing?",
is to develop {\bf search methods based on longer wavelengths}.

\noindent {\underline{Long-Wavelength Search Techniques}}

Several of these (which only utilize observations longward of 2\mic) 
are now proving successful:

\noindent
$\bullet$ {\it Rest-frame Optical Line Emission}

From the ground, the only currently practical method is to use deep imaging
in narrow-band filters (a 1\% filter width, for example makes a typical high-z
galaxy appear 0.5--1.5 magnitudes brighter than in a broad-band filter
at the same wavelength).
This method has been used successfully over the last 5 years (\eg, 
Malkan, Teplitz and McLean 1995).  
Figure 3 shows a ``color-magnitude" diagram for the field around 0953+549.
The 6 detected line-emitting galaxies (A through F) show significant
excesses in the 2.30\mic\ narrow-band filter relative to the broad one,
identified as \ha\ emission at z=2.49--2.50 (Malkan, Teplitz and McLean 1996;
Teplitz, Malkan and McLean 1998).

\begin{figure}[t]
\begin{center}
\epsfxsize=\textwidth\epsffile{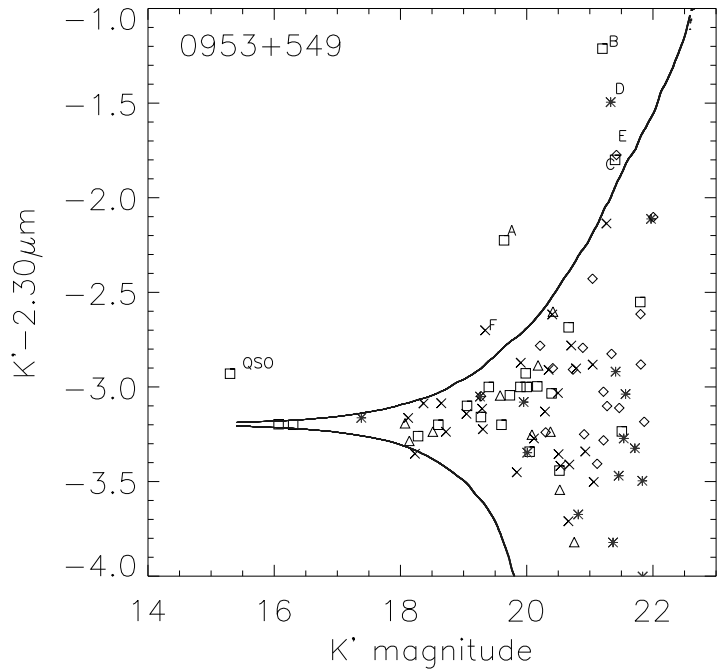}
\end{center}
\vspace{1.0truecm}
\caption{$\!\!$: Emission -Line Search with Narrow-band imaging:}
\smalltext{Vertical axis is magnitude excess brightness of each object in the
narrow-band (2.30\mic\ filter), with respect to the broad-band (K-short) filter.
The diverging curved lines show the $3\sigma$ error ranges.
There is a ``proto-cluster" of six galaxies which are 
significantly detected in emission.  If the line in the narrow filter is \ha\ as 
expected , it is at z=2.5. This has been confirmed for objects A and D, and is the
same as the redshift of the CIV absorption complex seen in the bright quasar in the
center of this field.}
\end{figure}

An extension of this method has been to search for the comparably strong
[OIII] 5007 line emission from star-forming galaxies at higher redshifts
(z=3--3.5, as opposed to 2--2.6 for redshifted \ha).  This has also detected
new $z \ge 3$ galaxies (Teplitz, Malkan, McLean 1999).

When NICMOS was--and will once again be--operating on HST, a unique method--slitless
spectroscopy (with the grisms, the G141 being most sensitive)--is available.
HST parallel observations of blank fields detected 33 emission-line galaxies this way.
The vast majority of these lines are believed to be \ha, at redshifts from
0.7 to 1.9 (Yan \et\ 1999).
The implied integrated star formation rates for galaxies at z=1.3 are
3 times higher than previous estimates based on {\it near}-UV continuum
(McCarthy \et 1999).
This discrepancy again implies substantial dust absorption of the far-UV continuum. 

\noindent
$\bullet$ {\it Damped \lya\ Absorption}

For decades, astronomers have searched for damped \lya\ absorption in the spectra
of bright background quasars to reveal HI clouds that are, or may become, young
galaxies.  The big question about these DLA systems is {\it how 
much star formation has actually taken place} 
at the redshifts we observe them.
In NICMOS coronographic imaging,
Colbert and Malkan (2000) found that none of 21 DLA systems at $z\sim 2$ is
plausibly identified with a near-infrared galaxy.
Figure 4 shows an example of a typical non-detection (the extra noise
under the 0.3-\arcsecs-radius coronographic hole in the upper left should be ignored, 
since it is not real).  The typical $5\sigma$ upper limits to the total galaxy
magnitudes that could have been detected in these 750--1000 second exposures
are H=21.5--22.5.
These upper limits to the brightness
at 1.6\mic\ derived for the 21 absorbers,
and are shown in Figure 5.  The solid line shows the magnitude that a DLA
galaxy would have at those redshifts, 
if it were going to evolve to a present-day $L_{*}$ galaxy.
These limits require that several of the DLA systems are more than 1 magnitude
fainter than an $L_{*}$ galaxy, while over half are fainter than $L_{*}$.
Hardly any of the 21 could be as bright as a half magnitude above $L_{*}$.

\begin{figure}[t]
\begin{center}
\epsfxsize=\textwidth\epsffile{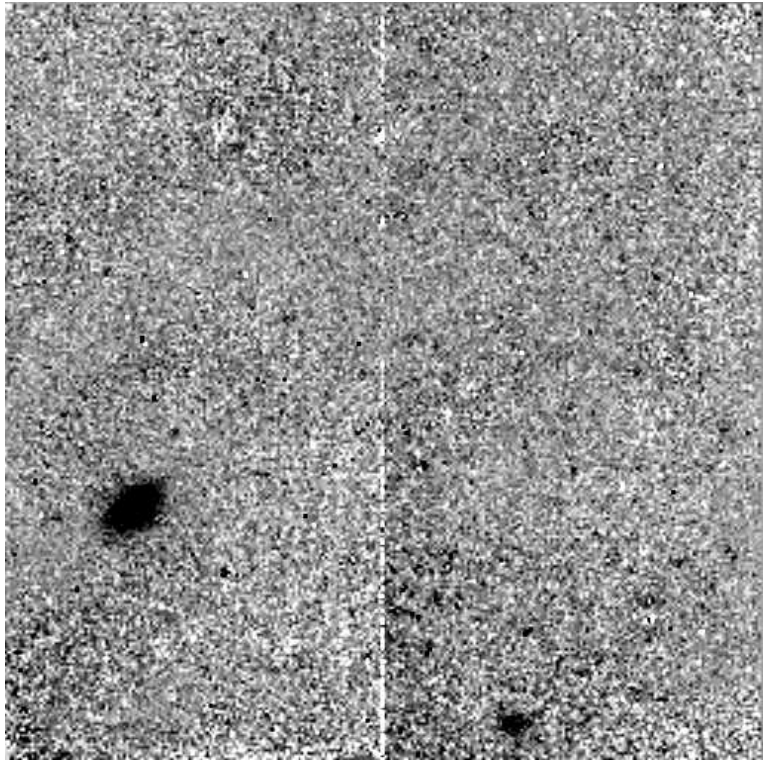}
\end{center}
\vspace{2.5truecm}
\caption{$\!\!$:NICMOS Coronographic Image of DLA Field }
\smalltext{The coronographic spot is in the upper-left hand quadrant.
This deep F160W image shows no plausible counterpart to the absorber,
since the two galaxies in the field are either unreasonably far away or too
bright.}
\end{figure}

\begin{figure}[t]
\begin{center}
\epsfxsize=\textwidth\epsffile{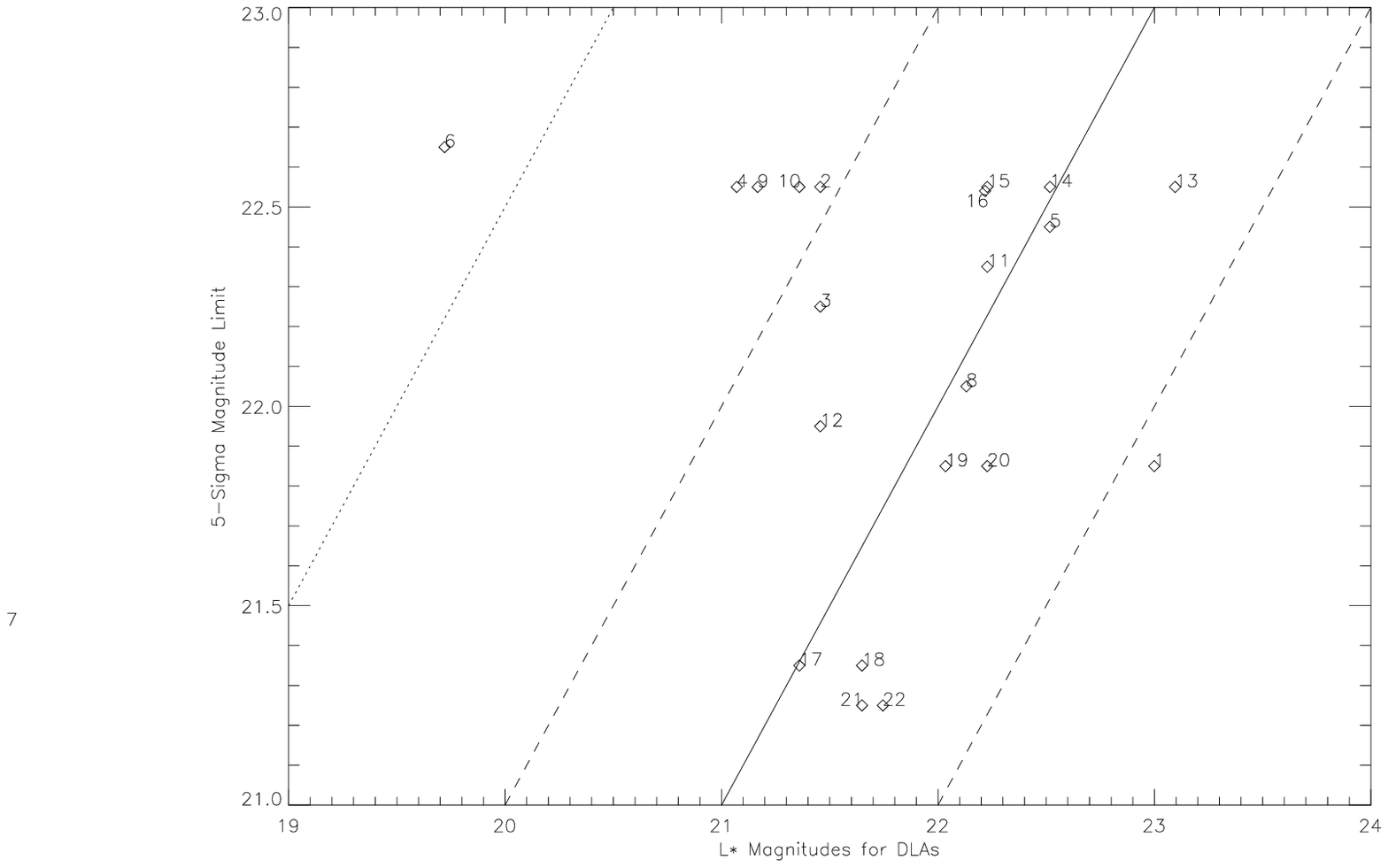}
\end{center}
\vspace{1.3truecm}
\caption{$\!\!$: NICMOS Coronographic Brightness Limits for DLAs, from Colbert and Malkan (2000) }
\smalltext{}
\end{figure}

\noindent
$\bullet$ {\it Deep Mid- and Far-infrared Imaging}

Another simple search method is to obtain ultra-deep imaging of blank fields
at mid-IR (7--25\mic), and far-IR (60--175\mic) wavelengths, and look for sources
with very large IR/optical flux ratios.  This has great promise when many square
degrees of sky can be covered at high sensitivity with cryogenic telescopes in
space (such as SIRTF). 
Rowan-Robinson \et\ (1997) has already used ISOCAM to find 
a significant number of 7 and 15\mic\
emitters in the Hubble Deep Field which emit as much or more power in the infrared
as in the optical.  It is likely that these are part of a substantial population
of dusty galaxies at redshifts of 1 or more.
Deep blank-field surveys using ISOPHOT at 175\mic, have found a large number of faint
sources with similarly high IR/optical ratios
(Kawara \et\ 1998; Puget \et\ 1999).

\noindent
$\bullet$ {\it Deep 850\mic\ Imaging with SCUBA}

At 850\mic, another breakthrough occurred when SCUBA provided the first
(small) array at that wavelength, 
so that spatially multiplexed observations could be made down to
flux levels of a few mJy (Smail, Ivison and Blain 1997). 
A surprisingly high surface density of faint red sources is being found, 
many of which are likely to have redshifts above 1 (Barger, Cowie and Richards 2000).

\smallskip
\smallskip
\smallskip
\smallskip

Many of the high-redshift galaxies {\it detected} in
the ground-based infrared narrow-band imaging searches 
to date were found in targeted regions
of the sky.  Since they were found in regions selected to contain one or more
high-redshift objects (quasars or absorption-line systems),
they might not serve as typical blank fields for deriving global statistics
(e.g. luminosity functions) for high-redshift galaxies.  It is possible that the
regions where some of these objects were found are biased to high density and
are therefore not representative of the Universe as a whole.  This issue will be
settled as the new generation of large-field infrared array detectors comes to
large telescopes.   

The remaining searches have already covered enough {\it un-}targeted blank sky
to indicate that long-wavelength observations are indeed picking
up a substantial population of galaxies too red for the optical photometry method.

\section{The Challenge of Spectroscopic and Far-IR Follow-up}

Aside from the relatively slow pace at which all of these long-wavelength methods currently
cover new sky, they also suffer from another difficulty:
how can a positive spectroscopic redshift be confirmed for the candidate
high-redshift galaxies?
Note that some of these considerations also apply to galaxies which produce
a significant fraction of their emission through an active nucleus
(presumably powered by accretion rather than stars).

In some cases, optical spectroscopy still works.
Several of the NICMOS Parallel Grism emission line galaxies have shown [OII] emission
lines at the same redshift indicated by their \ha\ lines in the near-IR
(Malkan \et 2000b).
A small number of the z=2.5 \ha-emitters 
have optical spectroscopic confirmations (Malkan, Teplitz, and McLean (1996);
Malkan \et (2000b), and one of them even turned out to be a background z=3.31 galaxy
which had been detected in the 2.16\mic\ filter by its 
redshifted [OIII] emission doublet.
However optical spectroscopy is not a general solution, because it relies on the same 
rest-frame far-UV that we already suspect is being absorbed by dust.

Many of these objects will not be spectroscopically confirmed and studied in detail
without near-IR spectroscopy.
Fortunately, the new generation of NIR spectrographs on large telescopes is now
capable of doing the job.  For example, NIRSPEC has confirmed the MTM ``Galaxy D" 
in the 0953+549 supercluster field, as it also shows the [OIII] doublet at
z=2.49, the same as its \ha\ line in the infrared (Malkan \et, 2000b).

Finally, if we do not require specific redshifts for individual galaxies,
we can still learn a significant amount about the Immature Universe
through the cruder measurements of ultradeep source counts and
even diffuse background measurements at long wavelengths.
As has often been noted, all the methods based on measurements at really long
wavelengths (beyond 100\mic) benefit from the added power of positive
rather than negative K-corrections. 
Somewhat coincidentally, they also suffer from the problem that the brightnesses
of galaxies at wavelengths longward of 100\mic\ are 
not nearly so well known as at shorter wavelengths, because IRAS only
observed out to 100\mic.
One approach has been to make theoretical calculations of the dust re-radiation
spectrum for various distributions of stars and interstellar matter
(Charlot and Fall 2000.)
The other approach is to assume that high-redshift galaxies are just the 
same as low-redshift ultraluminous galaxies of the same given luminosity,  
and use new ISOPHOT observations of them which extend out to 200\mic.
The first empirical description of the systematic tendency for more luminous galaxies
to have sharper sub-millimeter turnovers (since their strong star formation gives them
hotter far-IR peaks) was presented in Spinoglio \et\ (1995).
An improved version of this correlation, based on new ISOPHOT photometry, has been
made by Spinoglio \& Malkan (2000).

Several investigators have predicted the Diffuse Infrared Background (DIRB) spectrum.
The least theoretical (ie. most closely based on galaxy data) are the  ``backwards
evolution" models.  These start from a present-day luminosity function and galaxy
spectra, and assume some simple evolutionary law.
Figure 6 compares the Malkan \& Stecker 
(1998) model predictions with DIRB measurements which were made
{\bf subsequently}.  We found that the AGN contribution was only about a fifth
that of the non-Seyfert galaxies.

\begin{figure}[t]
\begin{center}
\epsfxsize=\textwidth\epsffile{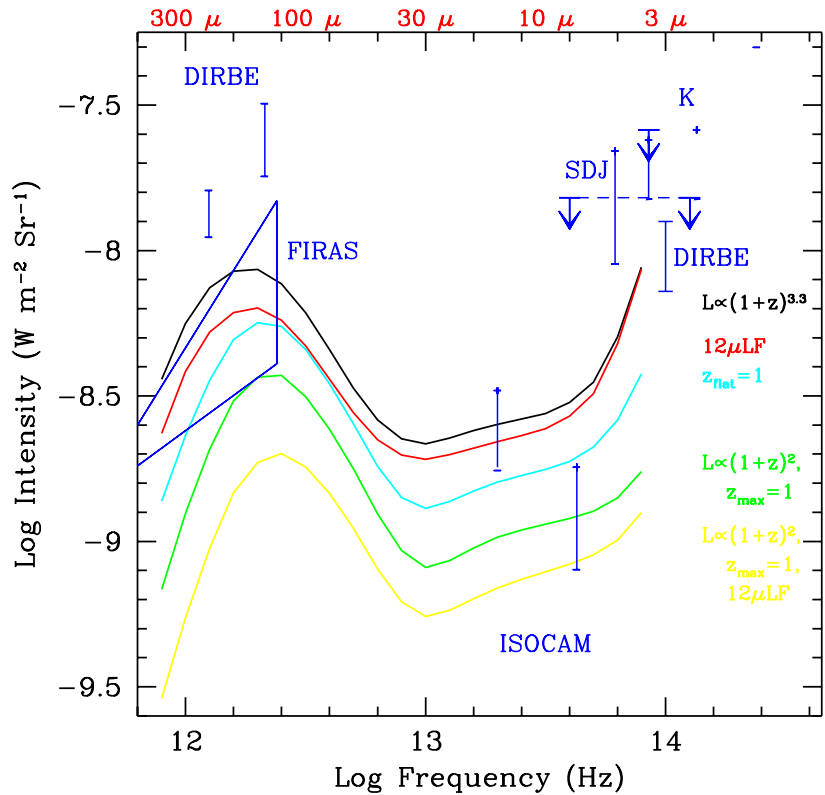}
\end{center}
\vspace{1.6truecm}
\caption{$\!\!$: Predictions from the Malkan and Stecker (1998) empirical model of the
Diffuse Infrared Background radiation. }
\smalltext{
The error bars show ISOCAM lower limits, and 
COBE detections made after the predictions were published, which are in good
agreement with the best model shown by the upper line.  It assumed that
galaxies--starting with the present-day luminosity function--evolved back in
time with luminosity increasing proportional to $(1+z)^{3.1}$, up to a redshift
of $z_{flat}=2$. Very similar results are obtained when the local LF is
determined from IRAS observations at either 60 or 12\mic.}
\end{figure}

\begin{figure}[t]
\begin{center}
\epsfxsize=\textwidth\epsffile{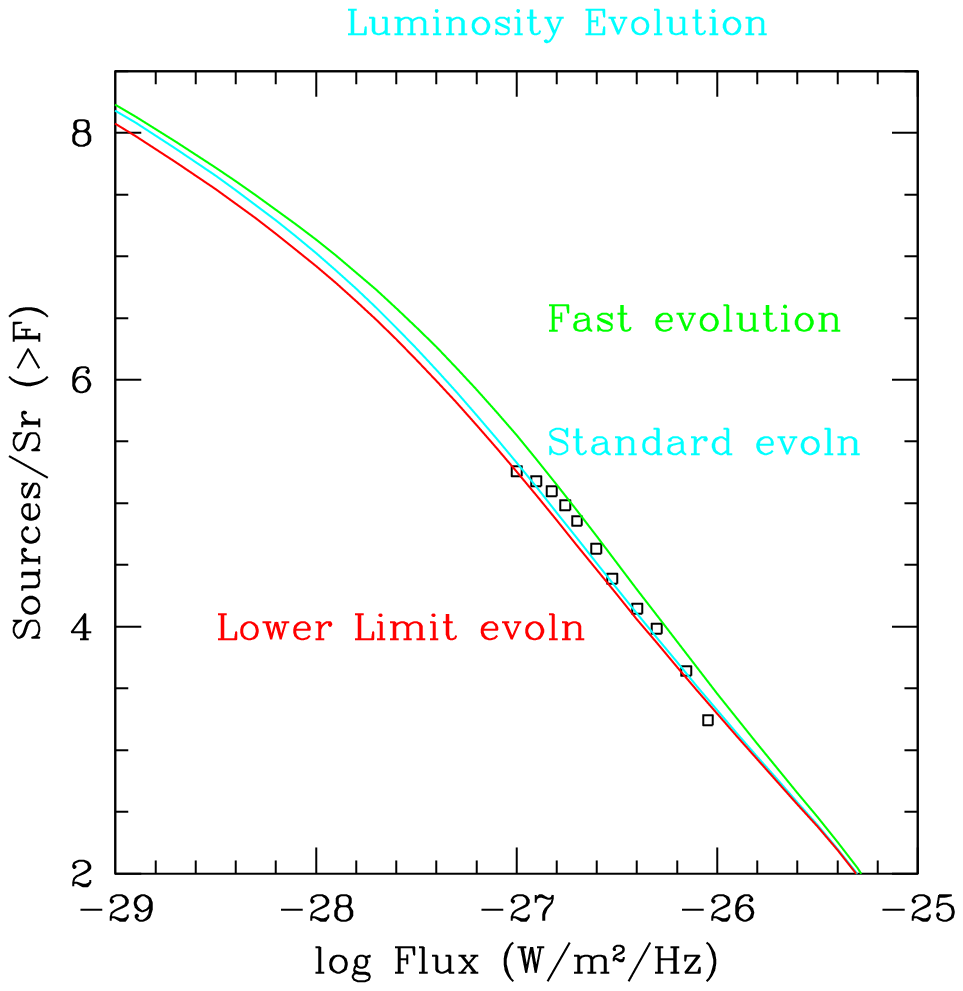}
\end{center}
\vspace{1.1truecm}
\caption{$\!\!$: Predictions of three Malkan and Stecker (1998) models for the deep source
counts at 175\mic.}
\smalltext{ 
The standard model (which happens to fit the observations
very well) assumes galaxy luminosity evolution going
as $(1+z)^{3.1}$ back to a redshift of $z_{flat}=2$. The ``fast evolution" model
has luminosity evolving as $(1+z)^{4.1}$, but only back to $z_{flat}=1.0$.
The most conservative ``lower limit" model assumes luminosity evolution as
$(1+z)^{3.1}$  back to $z_{flat}=1.3$.  All models then include galaxies
with a constant luminosity function from $z_{flat}$ back to z=5, but this
contribution does not add much to the counts.}
\end{figure}

Figure 7 uses the MS98 model to predict deep source counts at 175\mic.  
Again the agreement with later observations is excellent in both cases
(Malkan and Stecker 2000).
These successful {\it pre}dictions (not {\it post}dictions) 
suggest that the present day galaxy population has evolved its infrared properties
(at least since z=1--2) in a simple way, with a much higher number of high-luminosity
galaxies, whose bolometric power output is {\it primarily} at 25--100\mic\ rest
wavelengths.  (For example, at z=1.5 in these models, the predecessor of a modern
$L_*$ galaxy resembles a typical ``Ultraluminous Infrared Galaxy" from the IRAS 
survey.  These do not need to be considered a ``new population", but rather a
strongly evolving population whose remnants are still around today.)

\section {Conclusions}

Observational cosmologists have realized for decades that long-wavelength data would
eventually play a crucial role in studying representative regions of
the high-redshift Universe. (The power of
optical photometry and spectroscopy has turned out to be a less anticipated, pleasant
surprise.) This is based on the rough guide that, for a given sensitivity and number of
pixels, the longer
the observing wavelength, the better we can study all high-redshift 
galaxies.   We are now seeing this expectation realized across a wide front, 
from ground-based observations in the (sub-)millimeter and near-IR wavebands, to
space-based observations in the mid- and far-IR wavebands in between. 
The central goal of the next several years of hard work will be obtaining better 
spectral information--from spectroscopy, or from clever choice of imaging filters.
These observational advances (e.g. with SIRTF) will set the stage for the expected
breakthroughs with NGST.  Especially if NGST includes a mid- to far-IR capability,
we should be able to understand the first stages in the lives of galaxies as well as
we understand them in the present epoch.

\smallskip
\smallskip
It is a pleasure to thank Harry Teplitz for help in preparing some of the Figures,
and editorial improvements.


\begin{thebibliography}{}
\footnotesize
\parskip 0pt


\bibitem{} Barger, \et\ 2000, astro-ph 0001096

\bibitem{} Bechtold, J.; Elston, R.; Yee, H. K. C.; Ellingson, E.; Cutri, R. M. 1998
in The Young Universe: Galaxy Formation and Evolution at Intermediate and High Redshift. 
Edited by S. D'Odorico, A. Fontana, and E. Giallongo. ASP Conference
                Series; Vol. 146; 1998, p.241

\bibitem{} Charlot, S. \& Fall, S.M. 2000, astro-ph 0003128

\bibitem{} Colbert, J. \et\ 2000, in preparation

\bibitem{} Colbert, J. \& Malkan, M.A. 2000, in preparation

\bibitem{} Kawara, \et\ 1998 A\&A 336, L9

\bibitem{} Malkan, M.A. \et\ 2000a, in preparation

\bibitem{} Malkan, M.A. \et\ 2000b, in preparation

\bibitem{} Malkan, M.A. \& Stecker, F. S. 2000, in preparation

\bibitem{} Malkan, M.A. \& Stecker, F. S. 1998, ApJ, 496, 13

\bibitem{} Malkan, M.A., Teplitz, H. I., \& McLean, I.S. 1995 ApJL, 448, L5

\bibitem{} Malkan, M.A., Teplitz, H. I., \& McLean, I.S. 1996 ApJL, 468, L9

\bibitem{} McCarthy, P. \et\ 1999, ApJ 520, 548

\bibitem{} Pettini, M., Kellogg, M., Steidel, C.C., Dickinson, M., Adelberger, K.L., \& 
  Giavalisco, M., 1998, ApJ 508, 539

\bibitem{} Pettini, M. \et\ 2000, in preparation

\bibitem{} Puget, J. \et\  1999 A\&A 345, 29

\bibitem{} Rowan-Robinson, M. \et\ 1997, MNRAS 289, 490.

\bibitem{} Smail, I., Ivison, R.J., \& Blain, A.W. 1997, ApJL, 490, L5

\bibitem{} Spinoglio, L. \et\ 1995 ApJ, 453, 616

\bibitem{} Spinoglio, L. \& Malkan, M. 2000, in preparation

\bibitem{} Steidel, C. \et\ 1996, AJ 112, 352

\bibitem{} Steidel, C. \et\ 1998, ApJ 492, 428

\bibitem{} Teplitz,H.I, Malkan,M.A., \& McLean,I.S., 1999, ApJ 514, 33

\bibitem{} Teplitz,H.I, Malkan,M.A., \& McLean,I.S., 1998, ApJ 506, 519 

\bibitem{} Thompson, R.I. \et\ 1999 AJ, 117, 17

\bibitem{} Yan, L. \et\ 1999 ApJ Lett 519, 47.

\end{thebibliography}
\end{document}